\documentclass[letter,scriptaddress,twocolumn,prl,showkeys]{revtex4}

	\usepackage{amsmath}
	\usepackage{amsfonts}
  \usepackage{amssymb}
	\usepackage{makeidx}
	\usepackage{amsfonts}
	\usepackage[ansinew]{}
	\usepackage[usenames,dvipsnames]{pstricks}
	\usepackage{subfigure}
	\usepackage{epsfig}
	\usepackage{pst-grad} 
	\usepackage{pst-plot} 

\newcommand{\be}{\begin{equation}}
\newcommand{\ee}{\end{equation}}
\newcommand{\bea}{\begin{eqnarray}}
\newcommand{\eea}{\end{eqnarray}}

\makeindex

\begin{document}

\title{Directional dependence of the local estimation of $H_0$ and the non perturbative effects of primordial curvature perturbations}
\author{Antonio Enea Romano$^{1,2,3,4}$,Sergio Andr\'es Vallejo$^{2,4}$}

\affiliation{
${}^{1}$Yukawa Institute for Theoretical Physics, Kyoto University, Kyoto 606-8502, Japan;\\
${}^{2}$Department of Physics and CCTP, University of Crete, Heraklion 711 10, Greece \\
${}^{3}$Department of Physics, McGill University, Montr\'eal, QC H3A 2T8, Canada \\
${}^{4}$Instituto de Fisica, Universidad de Antioquia, A.A.1226, Medellin, Colombia\\
}

\begin{abstract}
Recent measurements  of the cosmic microwave background (CMB) radiation have shown an apparent tension with the present value of the Hubble parameter inferred from local observations of supernovae, which look closer, i.e. brighter, than what is expected in a homogeneous model with a value of $H_0$ equal to the one estimated from CMB observations. We examine the possibility that such a discrepancy is the consequence of the presence of a local inhomogeneity seeded by primordial curvature perturbations, finding that a negative peak of the order of less than two standard deviations could allow to fit low red-shift supernovae observations without the need of using a value of the Hubble parameter different from $H_0^{CMB}$. The type of inhomogeneity we consider does not modify the distance to the last scattering, making it compatible with the constraints of the PLANCK mission data. The effect on the luminosity distance  is in fact localized around the region in space where the transition between different values of the curvature perturbations occurs, producing a local decrease, while the distance outside the inhomogeneity is not affected. 

Our calculation is fully relativistic and non perturbative, and for this reason shows important effects which were missed in the previous investigations using relativistic perturbations or Newtonian approximations, because the structures seeded by primordial curvature perturbations can be today highly non linear, and relativist Doppler terms cannot be neglected. Because of these effects the correction to the luminosity distance necessary to explain observations is associated to a compensated structure which involves both an underdense central region and  an overdense outer shell, ensuring that the distance to the last scattering surface is  unaffected. 

Comparison with studies of local structure based on  galaxy surveys and luminosity density analysis reveals that the density profile we find could in fact be compatible with the one obtained for the same region of sky where is located most of \textit{the Cepheids used to calibrate the luminosity distance of the} supernovae employed for the local $H_0$ estimation, suggesting a possible directional dependence and \textit{calibration bias} which could be partially attributed to the presence of the Sloan Great Wall and hinting to the need of a more careful investigation, including a wider set of  Cepheids and low redshift supernovae in different regions of the sky.

\end{abstract}


\maketitle
\noindent\paragraph{\bf\emph{Introduction}}
Recent observations \cite{Ade:2013zuv} of the cosmic microwave background radiation (CMB) have pointed to an apparent  discrepancy between the value of the Hubble parameter $H_0$ deduced from cosmological data and the value inferred form local astrophysical observations \cite{Riess:2011yx,Ade:2013zuv,Verde:2013wza}.

Following \cite{Romano:2010nc} we denote with ${H_0}^{app}$  the value of the Hubble constant estimated from observational data ignoring the presence of an inhomogeneity and with  ${H_0}^{true}$ the value obtained taking into account the inhomogeneity.
According to this notation, since both the above mentioned estimations were based on the assumption of homogeneity, the apparent tension can be expressed as $H_{0,SN}^{app}\approx1.09H_{0,CMB}^{app}$, where $H_{0,SN}^{app}$ and $H_{0,CMB}^{app}$ are the values estimated from fitting respectively low-redshift supernovae and CMB observations.
One possible explanation of such a difference could be the effects of the local structure on the analysis of cosmological and astrophysical observations, and we will show what kind of inhomogeneity could resolve the tension, i.e. could give 
$H_{0,SN}^{true}=H_{0,CMB}^{true}$.

The effects of a local inhomogeneity on the estimation of cosmological parameters have been studied already in different contexts such as the estimation of the equation of state of dark energy \cite{Romano:2010nc}, where it has been shown that ignoring the presence of local inhomogeneities in data analysis can lead to the wrong conclusion of an apparent evolving dark energy, while only a cosmological constant is in fact present, or to corrections \cite{Romano:2012gk} to the apparent value of the cosmological constant.
More recently \cite{Romano:2013kua} it was shown that a present day local inhomogeneity seeded by a peak of primordial curvature perturbations could in fact not only affect the estimation of the value of the cosmological constant, but also of $H_0$. 
In particular it was shown that the  effects of such an inhomogeneity on $H_0$ depend on the spatial gradient of the inhomogeneity, and could be important independently from the amplitudes of the curvature perturbations. This implies that this kind of inhomogeneities arise naturally from fluctuations of the primordial curvature perturbations, and require a careful investigation.
In this letter we will focus on the early time origin of an inhomogeneity able to solve the apparent tension in the $H_0$ estimation, showing how the inflationary scenario can easily explain the formation of such a local structure today. 

Some of the previous studies of the effects of inhomogeneities on the expansion rate, such as for example \cite{Wojtak:2013gda,Wang:1997tp,Shi:1997aa},  were based on an implicit assumption of the validity of the Hubble law and on the use of a  Newtonian approximation to relate the local velocity field to $H$. Other approaches are  based on the Hubble bubble model \cite{Marra:2013rba}, which is the basis of the top hat spherical collapse, or on the averaging of the perturbations \cite{Ben-Dayan:2014swa}.
All these different attempts to estimate the effects of inhomogeneities are based on a Newtonian or relativistic linear perturbations approximation, and as such are not able to obtain the non perturbative relativistic effects that the use of an exact solution of Einstein's equations allows to calculate.
In fact, as shown in \cite{Bolejko:2012uj}, perturbation theory is not able to fully account for the relativistic effects of an inhomogeneities for the luminosity distance, because gradient terms can have important contributions which are normally ignored in the Newtonian or relativistic perturbative calculations. 

The discrepancy between apparent and true values of cosmological parameters is a consequence of the intrinsic limitation of cosmological or astrophysical observations involving redshift measurements, implying an intrinsic uncertainty \cite{Romano:2012tt} on  the estimation of cosmological parameters.
Under the assumption of homogeneity the redshift is explained exclusively as a consequence of the expansion of the Universe, while taking into account spatial inhomogeneity additional contributions to the redshift can come from the spatial variation the gravitational potential. While the effects of these variations are normally considered to be small compared to the contribution associated to the Universe expansion, it is possible that the local structure can actually induce some important corrections respect to the apparent values, i.e. the ones obtained from observation ignoring the space variation of the metric.

From the relation between the Hubble parameter and the luminosity distance in a spatially flat FLRW Universe:
\bea
D_L(z)&=&(1+z)\int^{z}_0\frac{d x}{H(x)}\,, \\
H^{\Lambda CDM}(z)&=&H_0^{app}\sqrt{\Omega_M(1+z)^3+\Omega_{\Lambda}}\,,
\eea
we can see that the larger estimated value of $H_{0,SN}^{app}$ compared to $H_{0,CMB}^{app}$ corresponds to the observation of supernovae closer than what is expected in a $\Lambda CDM$ model with the Hubble parameter equal to $H_{0,CMB}^{app}$.
In the framework of homogenous cosmological models the observed brightness of low-redshift supernovae is interpreted as the result of 
a faster expansion of the Universe, while if the effects of a local inhomogeneity are properly taken into account there is no need to invoke a value of $H_0$ different from $H_{0,CMB}^{app}$.
In the particular case of the set of observations we are interested in this letter we have that an appropriate local inhomogeneity does not affect the distance to the last scattering, and consequently the estimation of $H_0$ from CMB data, so that the assumption of homogeneity in analyzing CMB data does not give a different result respect to the corresponding data analysis taking into account the inhomogeneity, i.e. $H^{app}_{0,CMB}\approx H^{true}_{0,CMB}$.
The presence of a local inhomogeneity could instead affect the local estimation of $H_0$, so that $H^{app}_{0,SN}>H^{true}_{0,SN}$, where $H^{true}_{0,SN}$ is the value of $H_0$ which according to the above definitions is obtained from the data  taking into account the effects of the inhomogeneity. 
 In presence of an inhomogeneity in fact the determination of $H_0$ from $D_L^{obs}(z_{SN})$, where $z_{SN}\approx 0.04$ is the mean redshift of the supernovae observations used for the local estimation of $H_0$ \cite{Riess:2011yx}, should not be based on  the standard $\Lambda CDM$ redshift distance relation, but another one which takes into account the local inhomogeneity, and in this way the discrepancy can be solved, i.e. $H^{app}_{0,CMB}\approx H^{true}_{0,CMB}=H^{true}_{0,SN}$.\\

\noindent\paragraph{\bf\emph{Low red-shift supernovae and the $H_0$ estimation}}
We will adopt the following notation for the luminosity distance, $D_L^{mod}(z,H_0)$, where we are denoting with a superscript the cosmological model or the observed data, and $H_0$ is explicitly treated as an argument to avoid confusion.
According to the above notation low red-shift supernovae observations are fitted by a homogeneous cosmological model according to
\be
D_L^{hom}(z_{SN},H_{0,SN}^{app})=D_L^{obs}(z_{SN})\,,
\ee
with $H_{0,SN}^{app}\approx 1.09 H_{0,CMB}^{app}$.
We will show that if the effects of primordial curvature perturbations are taken into account
a local inhomogeneity could resolve the apparent discrepancy, i.e.
\be
D_L^{inh}(z_{SN},H_{0^,CMB}^{true})=D_L^{obs}(z_{SN})\,.
\ee

Such an inhomogeneity can arise naturally from primordial curvature perturbations as predicted by inflation and constrained by CMB observations to have a standard deviation of about $5 \times 10^{-5}$, providing a simple mechanism for its origin.\\
\noindent\paragraph{\bf\emph{Modeling the local Universe}}
We will model the local structure with a LTB solution of Einstein's equation with a cosmological constant term, assuming to be located around its center. This is a pressureless spherically symmetric solution which allows to take into account the non perturbative  effects due to the inhomogeneity. The central location assumption can be interpreted \cite{Romano:2011mx} as calculating the monopole contribution to the corrections coming from the local structure.
Since we are considering inhomogeneities which can be seeded by not very large fluctuations of primordial curvature perturbations,  these structures can arise very naturally.
Only the use of an exact solution allows to study their formation, since today they can become highly non linear, and the standard theory of structure formation based on a perturbative approach cannot be applied to study their full evolution.

The LTB solution is given by
 \cite{Lemaitre:1933qe,Tolman:1934za,Bondi:1947av} 
\begin{eqnarray}
\label{LTBmetric} %
ds^2 = -dt^2  + \frac{\left(R,_{r}\right)^2 dr^2}{1 + 2\,E(r)}+R^2
d\Omega^2 \, ,
\end{eqnarray}
where $R$ is a function of the time coordinate $t$ and the radial
coordinate $r$, $E(r)$ is an arbitrary function of $r$, and
$R_{,r}=\partial_rR(t,r)$.
The Einstein's equations with a cosmological constant give
\begin{eqnarray}
\label{eq2} \left({\frac{\dot{R}}{R}}\right)^2&=&\frac{2
E(r)}{R^2}+\frac{2M(r)}{R^3}+\frac{\Lambda}{3} \, , \\
\label{eq3} \rho(t,r)&=&\frac{2M,_{r}}{R^2 R,_{r}} \, ,
\end{eqnarray}
where $M(r)$ is an arbitrary function of $r$, $\dot{R}=\partial_t R(t,r)$ and $c=8\pi G=1$ and is assumed in the rest of the
paper. We will also adopt, without loss of generality, the coordinate system in which $M(r)\propto r^3$,  fix the geometry of the solution by using a function $k(r)$ according to $2E(r)=-k(r)r^2$, and consider models with a vanishing bang function, i.e.  $t_b(r)=0$.\\
\noindent\paragraph{\bf\emph{The effects of primordial curvature perturbations}}
The metric after inflation near a peak of the primordial curvature perturbation \cite{Romano:2010nc,Romano:2013kua} can be written as
\be
ds^2=-dt^2 + a_F^2(t)e^{2\zeta({ r})}(dr^2+r^2d\Omega^2)\,, \label{minf} 
\ee
where $\zeta(r)$ is the primordial curvature perturbation.
The spherical symmetry approximation is justified by the  property of a
Gaussian random field \cite{Bardeen:1985tr} according to which
large peaks of a stochastic function tend to have a spherical
shape. 
At a sufficiently early time when both the LTB and the metric in eq.(\ref{minf}) are valid, they can be matched \cite{Romano:2010nc} leading to the following relation
\begin{equation}
 k(r)=-\frac{1}{ r^2}[(1+r\zeta')^2-1] \,. \label{kz}
\end{equation}
The use of the LTB solution allows to study exactly the formation of the structure seeded by $\zeta(r)$, and in particular to go beyond the limitations of perturbation theory. It turns out in fact that such early time tiny curvature perturbations of the order of $10^{-5}$ can lead to a highly non linear structure today whose effects cannot be neglected.

Eq.(\ref{kz}) is very important because it gives the connection between the early and the present Universe, and it can be used in two ways: given an early Universe curvature perturbations it allows to find the present day structure seeded by it, or vice versa, given a present day structure it can be used to find what primordial curvature perturbation could have originated it. Normally the first approach is adopted\cite{Romano:2010nc,Romano:2013kua}, but in this letter we will adopt the second one, since it is more convenient to define the model of the inhomogeneity  in terms of $k(r)$.
In this way we can  find a natural explanation for the origin of the present day highly non linear structure whose effect cannot be captured by perturbations theory, while the use of the LTB solution gives the non perturbative evolution of the structure.\\

\noindent\paragraph{\bf\emph{Effects on the luminosity distance}}
To compute $D_L^{\Lambda LTB}(z)$ we need to solve the radial null geodesics 
\begin{eqnarray}
{dr\over dz}&=&{\sqrt{1+2E(r(z))}\over {(1+z)\dot {R'}[T(r(z)),r(z)]}}  \,,
\label{eq:34} \\
{dt\over dz}&=&-\,{R'[T(r(z),r(z))]\over {(1+z)\dot {R'}[T(r(z)),r(z)]}} \,,
\label{eq:35} 
\end{eqnarray}
and then we substitute in the formula for the luminosity distance in a LTB space
\be
D_L^{inh}(z,H_{0}^{true})=D^{\Lambda LTB}(z)=(1+z)^2 R(t(z),r(z)) \,
\ee
where we set the parameters of the LTB solution so that 
\be
H_0^{LTB}=\frac{2}{3}\frac{\dot{R}(t_0,0)}{R(t_0,0)}+\frac{1}{3}\frac{\dot{R}'(t_0,0)}{R'(t_0,0)}=H_{0}^{true}.
\ee
For homogeneous cosmological models we assume a flat $\Lambda CDM$ solution according to
\bea
H(z)&=&H_0^{app}\sqrt{\Omega_M(1+z)^3+\Omega_{\Lambda}}\,, \nonumber \\
D_L^{hom}(z,H_0^{app})&=&(1+z)\int^{z}_0\frac{d x}{H(x)}\,.
\eea
The above definitions are very important since they give an explicit mathematical definition of $H_0^{app}$ and $H_0^{true}$, which is coherent with the notion of apparent observable adopted in this paper and which can be expressed as
\bea
D_L^{hom}(z,H_0^{app})=D_L^{inh}(z,H_0^{true})=D_L^{obs}(z).
\eea
We consider solutions with the function $k(r)$ given by a gaussian
\bea
k(r)=A e^{-\left(\frac{r-r_0}{\sigma}\right)^2}\,,
\eea
centered  approximately at the radial comoving coordinate of low-redshift supernovae used for  the local estimation of $H_0$.
Different values of $r_0$ can be chosen to control the shape  of the inhomogeneity. 
This solution corresponds to a compensated inhomogeneity which is asymptotically homogeneous and, if the width of the gaussian is sufficiently small, there is no important difference between the central and asymptotic value of $k(r)$. This property is quite important because it ensures that the Universe is effectively a flat $\Lambda CDM$ at the center and sufficiently far from it, making sure that the effect on the luminosity distance at the last scattering is very small.

We will adopt a system of units in which $H_{0,CMB}^{true}=1$, and use the fiducial value $\Omega_{\Lambda}=0.692$.
The time used as initial condition at the center  for the geodesics equations is obtained by integrating the Einstein equation in the scale factor from the big-bang till today.

%

As shown in fig.(3), since the type of inhomogeneity we have considered is compensated, it has no effect on the luminosity distance outside it, i.e.
\be
D^{hom}(H_{0,CMB}^{true},z_{LS})\approx D^{inh}(H_{0,CMB}^{true},z_{LS})\,,
\ee
where $z_{LS}$ is the redshift of the last scattering surface, or any redshift sufficiently larger than $z_{SN}$. This is due to the fact that the central and the asymptotic value of the curvature function is approximately the same. 
According to our definitions of apparent observable, fitting $CMB$ data with a homogeneous or inhomogeneous model is consequently giving the same estimate for $H_{0,CMB}^{app}\approx H_{0,CMB}^{true}$.

For the case of low redshift supernovae instead it can be seen in fig.(2) that
\be D^{inh}(H_{0,CMB}^{true},z_{SN})\approx D^{hom}(H_{0,SN}^{app},z_{SN})=D^{obs}_{z_{SN}}\,,
\ee 
which implies that 
\be 
H_{0,SN}^{app}>H_{0,SN}^{true}=H_{0,CMB}^{true} \,.
\ee
This shows how the effects of the inhomogeneity are able to resolve the apparent disagreement between cosmological and local estimations of $H_0$, since the luminosity distance is modified only around $z_{SN}$.

As seen in fig.(2,3) the effect of the structure seeded by this primordial curvature perturbation profile corresponds to and underdense central region followed by an overdense shell, which is what we expect for a compensated inhomogeneity, due to the fact that the function $k(r)$ is asymptotically zero. 
The luminosity distance is decreased respect to the homogeneous case both in the under and over dense regions, a relativistic effect associated to large gradient terms which are normally neglected in Newtonian or perturbative calculations as shown for example in \cite{Bolejko:2012uj}.
This inhomogeneity is seeded by a curvature perturbation of less than two standard deviations 
as shown in  fig.(1), and the effect on the luminosity distance is associated to the region of transition between different values of $\zeta(r)$.
Such a perturbation can occur quite naturally according the inflationary scenario, and smaller amplitude perturbations could still induce some important effects.

The density profile shown in fig.(3) is compatible with the results of the analysis of galaxy surveys\cite{Keenan:2013mfa}.
Quite remarkably \textit{the Cepheids used to calibrate the luminosity distance of the} supernovae employed for the local estimation of $H_0$ in \cite{Riess:2011yx} are inside the region with the largest overdensity(Subregion 2 in  \cite{Keenan:2013mfa}) with a density profile compatible with our model in the region of available data, suggesting a possible \textit{calibration bias} and general directional dependence$^{1}$of the local structure effects on the $H_0$ estimation which could be partially attributed to the presence in the region of the Sloan Great Wall \cite{Gott:2003pf}. Since the calibration affects the analysis of all the supernovae in any direction, the presence of this inhomogeneity could significantly bias the estimation of $H_0$ based \cite{Riess:2011yx,Riess:2016jrr} on the analysis of all the low red-shift supernovae. Only a detailed analysis could confirm if the effects on $H_0$ are mostly due to the inhomogeneity in the angular region of the Cepheids, and the consequent bias on the calibration. It could be that inhomogeneities also have a strong effect on the luminosity distance of the supernovae located in directions different from the Cepheids, or that both effects are important.
Nevertheless the remarkably good qualitative agreement between the density profile of the model we propose to explains the $H_0$ discrepancy and independent luminosity density and galaxy surveys data  suggests that  calibration bias could be the main source of the $H_0$ estimation discrepancy.

\footnotetext[1]{\textit{The Cepheids used to calibrate the luminosity distance of the} supernovae are in the same right ascension range of Subregion 2, but span a wider declination range, so here we are assuming a similar density profile would be obtained if the galaxy survey analysis was extended to the same declination range of the Cepheids, or alternatively if $H_0$ was estimated only using the supernovae within the same declination range of Subregion 2.}
 
A careful investigation of such a directional dependence goes beyond the scope of this paper, so we will not pursue it here, but we expect our calculations to provide a good estimation of the effects due to the density profile along that particular direction. It would be interesting though to check carefully if using supernovae in different directions away from Sloan Great Wall the $H_0$ estimation could change significantly from the one obtained by \cite{Riess:2011yx}.

\begin{figure}
\label{krzr}
	\centering
	\begin{tabular}{c}
\includegraphics[scale=0.7]{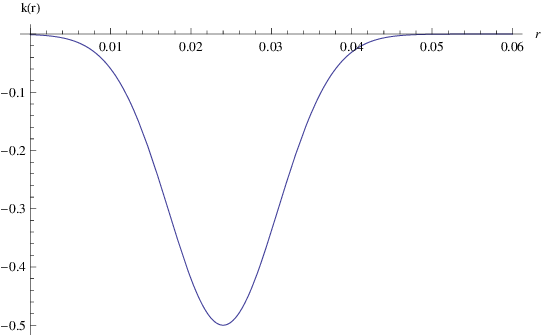} \\
\includegraphics[scale=0.8]{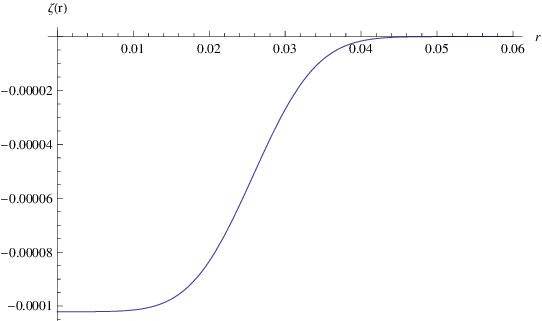}
\end{tabular}
\caption{The primordial curvature perturbation $\zeta(r)$ and the function $k(r)$
are plotted for $A=-0.5,r_0=r_{SN}*0.8,\sigma=r_0/2.5$, and $r_{SN}=r^{\Lambda CDM}$. The quantities $A$ and $k(r)$ are in units of $(H_{0,CMB}^{true})^2$, while $r_0$ is in units of $(H_{0,CMB}^{true})^{-1}$.The effect on the luminosity distance corresponds to the region of transition between different values of the curvature perturbation.} 

\end{figure}

\begin{figure}[h]
\label{dD}
		\centering
	\begin{tabular}{c}
		\includegraphics[scale=0.5]{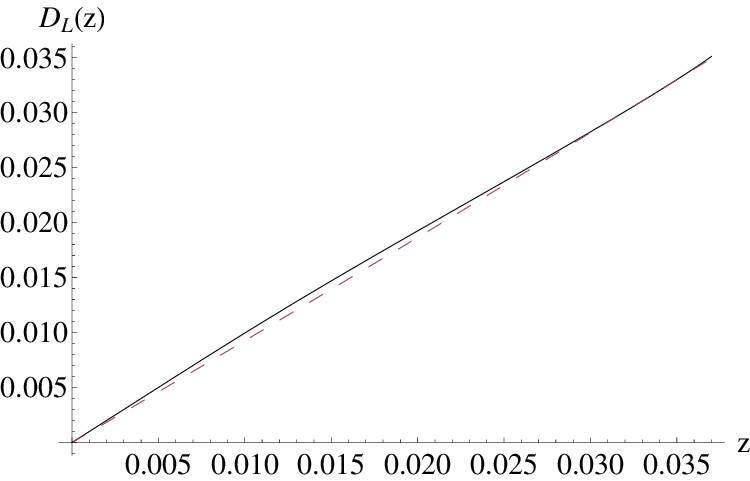}
	\end{tabular}
\caption{According to the definition in eq.(12-14), the luminosity distance $D^{hom}(H_{0,SN}^{app},z)$ for a homogeneous model  and the luminosity distance $D^{inh}(H_{0,CMB}^{true},z)$ for an inhomogeneous model are plotted in units of $(H_{0,CMB}^{app})^{-1}$respectively with a dashed and solid line. 
Despite the difference in the value of $H_0$ for the two models, $D^{inh}(H_{0,CMB}^{true},z_{SN})\approx D^{hom}(H_{0,SN}^{app},z_{SN})$, implying that the observed luminosity distance $D^{obs}_L(z_{SN})$ can be fitted using $H_{0,CMB}^{true}\approx H_{0,CMB}^{app} $ instead of $H_{0,SN}^{app}$ if the effect of the inhomogeneity is taken into account by using $D^{inh}$ instead of $D^{hom}$, resolving the apparent discrepancy between local and cosmological observations.}
\end{figure}

\begin{figure}
\label{drhodD2}
		\centering
	\begin{tabular}{c}
		\includegraphics[scale=0.8]{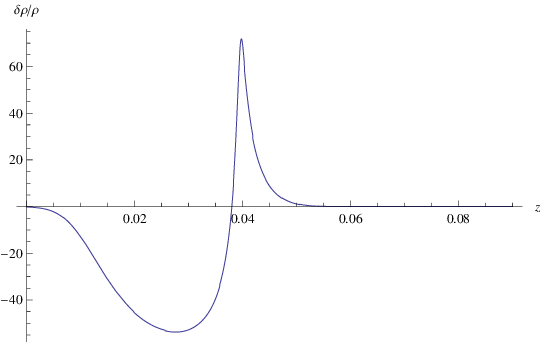} \\
		\includegraphics[scale=0.6]{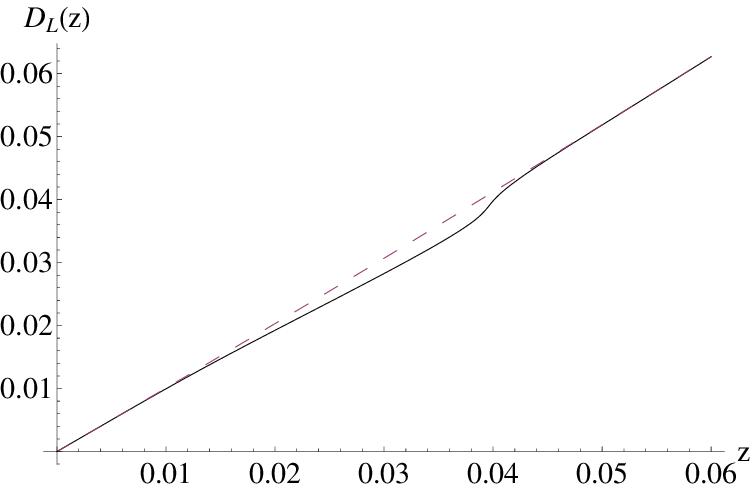}
	\end{tabular}
\caption{On the top the percentage density contrast $\frac{\delta\rho}{\rho}=100(1-\frac{\rho^{LTB}(z)}{\rho^{\Lambda CDM}(z)})$ is plotted as a function of the redshift, showing how contrary to the linear theory approximation, when non perturbative effects are taken into account, not only underdense regions but also overdense regions can be associated to the decrease of the luminosity distance necessary to explain observations. 
On the bottom we plot in units of $(H_{0,CMB}^{true})^{-1}$ the luminosity distance for a $\Lambda CDM$ model and an inhomogeneous model. In both the top and bottom plot the value of the Hubble parameter is the same for the homogeneous and inhomogeneous models, $H_0=H_{0,CMB}^{true}$.
The dashed line is the plot of $D^{hom}(H_{0,CMB}^{true},z)$ while the solid line is for $D^{inh}(H_{0,CMB}^{true},z)$. The luminosity distance is approximately the same outside the inhomogeneity, for $z>z_{SN}$, implying that the distance to the last scattering is not affected by the inhomogeneity, and consequently $H_{0,CMB}^{true}\approx H_{0,CMB}^{app}$. This shows that the analysis of the CMB data performed under the assumption of homogeneity is not introducing any misestimation for $H_0$, contrary to its local estimation, which is based on luminosity distance observations affected by it as shown in fig.(2).
}
\end{figure}

\noindent\paragraph{\bf\emph{Conclusions}}
We have shown how the apparent value of $H_{0,SN}^{app}$ obtained from  analyzing the supernovae low red-shift luminosity distance data under the assumption of homogeneity can receive an important correction when the local structure is taken into proper account. 
The apparent disagreement is due to the different way in which the $H_0$ estimation is affected by the presence of the local inhomogeneity we have considered. In the case of $CMB$, the inhomogeneity does not affect the distance to the last scattering surface, implying that analyzing $CMB$ data under the assumption of homogeneity does not introduce any misestimations.
For low red-shift luminosity distance instead there can be an important impact which could resolve the discrepancy, or at least could be able to account for part of the difference.
The type of inhomogeneity able to explain the difference can be seeded by a primordial curvature perturbation of the order of less than two standard deviations, and as such can arise quite naturally according to the inflationary scenario.
This shows the importance of non pertubative effects for any local observation in order to avoid this kind of apparent tension with other measurements less affected by the local structure.

Quite remarkably  the  comparison with galaxies redshifts surveys suggests that the \textit{the Cepheids used to calibrate the luminosity distance of the} supernovae employed for local estimation of $H_0$ are in a region of sky associated to a particularly high overdensity and with an inhomogeneity profile compatible with the one we have studied, which could be partially attributed to the presence in that region of the Sloan Great Wall.
Since the calibration affects the analysis of all the supernovae in any direction, the presence of this inhomogeneity could significantly bias the estimation of $H_0$ based \cite{Riess:2011yx,Riess:2016jrr} on the analysis of all the low red-shift supernovae. Only a detailed analysis could confirm if the effects on $H_0$ are mostly due to the inhomogeneity in the angular region of the Cepheids, and the consequent bias on the calibration. It could be that inhomogeneities also have a strong effect on the luminosity distance of the supernovae located in directions different from the Cepheids, or that both effects are important.
Nevertheless the qualitative agreement between the density profile of the model we propose to explains the $H_0$ discrepancy and independent luminosity density and galaxy surveys data  suggests that the calibration bias could be the main source of the discrepancy.

It would be interesting in the future to go beyond the monopole contribution and extend the study to the possible directional dependency of the local $H_0$ estimation. 
In this way it would be possible to determine indirectly the higher multipoles of the local structure which causes the apparent angular dependency of $H_0^{app}$. 
The same type of analysis could be extended to other cosmological parameters such as for example the effective equation of state of dark energy, which have already been shown to be affected by local inhomogeneities, but are normally analyzed under the assumption of isotropy.\\
  
\acknowledgments
AER thanks  Ryan Keenan and Licia Verde for useful comments and discussions, the Department of Physics of Heidelberg University for the kind  hospitality, and Luca Amendola, Valeria Pettorino and Miguel Zumalacarregui for insightful discussions. 
The authors thank the anonymous referee for useful comments to improve the the manuscript.
This work was supported by the European Union (European Social Fund, ESF) and Greek national funds under the ``ARISTEIA II'' Action, the Dedicacion exclusica and Sostenibilidad programs
at UDEA, the UDEA CODI project IN10219CE. 


\end{document}